\documentclass[twocolumn,amssymb,prl,epsfig,superscriptaddress]{revtex4}
\usepackage[utf8]{inputenc}
\usepackage{color}
\usepackage[normalem]{ulem}
 
\usepackage{amssymb}

\usepackage{graphicx}
\usepackage{epsfig}
\usepackage{amssymb}

\begin{document}

\title{Charge and Spin Density Waves observed through their spatial fluctuations   by coherent and simultaneous  X-ray diffraction}

\author{ V. L. R. Jacques}
\affiliation{Laboratoire de Physique des Solides, Universit\'e Paris-Sud, CNRS, UMR 8502, F-91405 Orsay, France}

\author{E. Pinsolle}
\affiliation{Laboratoire de Physique des Solides, Universit\'e Paris-Sud, CNRS, UMR 8502, F-91405 Orsay, France}

\author{S. Ravy}
\affiliation{Synchrotron SOLEIL - L'Orme des Merisiers Saint-Aubin - 91192 GIF-sur-YVETTE CEDEX, France}

\author{G. Abramovici}
\affiliation{Laboratoire de Physique des Solides, Universit\'e Paris-Sud, CNRS, UMR 8502, F-91405 Orsay, France}

\author{D. Le Bolloc'h}
\affiliation{Laboratoire de Physique des Solides, Universit\'e Paris-Sud, CNRS, UMR 8502, F-91405 Orsay, France}

\begin{abstract}

Spatial fluctuations of spin density wave (SDW) and charge density wave (CDW) in chromium have been compared by combining coherent and simultaneous X-ray diffraction experiments. Despite their close relationship, spatial fluctuations of the spin and of  the charge density waves display a very different behavior: the satellite reflection associated to the charge density displays speckles while the spin one displays an impressive long-range order. This observation is hardly compatible with the commonly accepted magneto-elastic origin of CDW in chromium and is more consistent with a purely electronic scenario where CDW is the second harmonic of SDW. A BCS model taking into account a second order nesting predicts correctly the existence of a CDW and explains why the CDW is more sensitive to punctual defects.

\end{abstract}

\maketitle

Studying systems in which two phases coexist is particularly interesting, not only 
to determine the coupling between two phases, but also to better understand
  the origin of each phase individually.  As examples, the coexistence of superconducting and charge density wave states
 in cuprates\cite{letacon}, or  in conventional
superconductors\cite{nbse2}, and of coexisting singlet and triplet
superconductivity\cite{GA} have recently been investigated. This paper is devoted to the study
of  coexisting  Charge Density Wave (CDW) and  Spin Density Wave (SDW).
  By which process is a
periodic modulation of charges connected to a periodic modulation of spin?
The answer lies in the coupling between the two phases,  which are
observed here through their spatial fluctuations, and not through their
average behavior, as is usually done. For this purpose
we have mixed two experimental techniques, coherent X-ray diffraction and
simultaneous diffraction, to probe a model system of itinerant
antiferromagnetism such as chromium.

Chromium is unique in its kind. It is the only transition metal stabilizing a SDW
state.  Despite the simplicity of its atomic structure (body centered cubic), it
stabilizes a complicated antiferromagnetic state made of a CDW and a SDW. The  wave
vectors associated to these modulations are collinear and incommensurate with the crystal
lattice at all temperatures\cite{hill}.  The period of the SDW is large, running over more than $1/\delta=21$ unit cells. The CDW is associated to a periodic
lattice distortion and its period is twice shorter.

The origin of SDW in chromium is now clear and arises from the peculiar geometry
 its Fermi surface\cite{fawcett,kevan}. It is based on a nesting effect between the
electron pocket centered at point $\Gamma$ and the hole pocket centered at the edge of the Brillouin zone at point  $H$ (see Fig. 1)\cite{overhauser}. Note that the two pockets do not have the same size and this size difference explains the presence of a SDW in chromium. 

 In contrast, the physical
origin of  the CDW  in chromium is still not understood. Two scenarios may be considered.
Either the CDW is induced by a magnetostriction effect (a coupling
between elasticity and magnetism) or by a purely electronic effect based on nesting between electronic bands\cite{youngsokoloff}.

\begin{figure}
\center
\includegraphics[width=8cm]{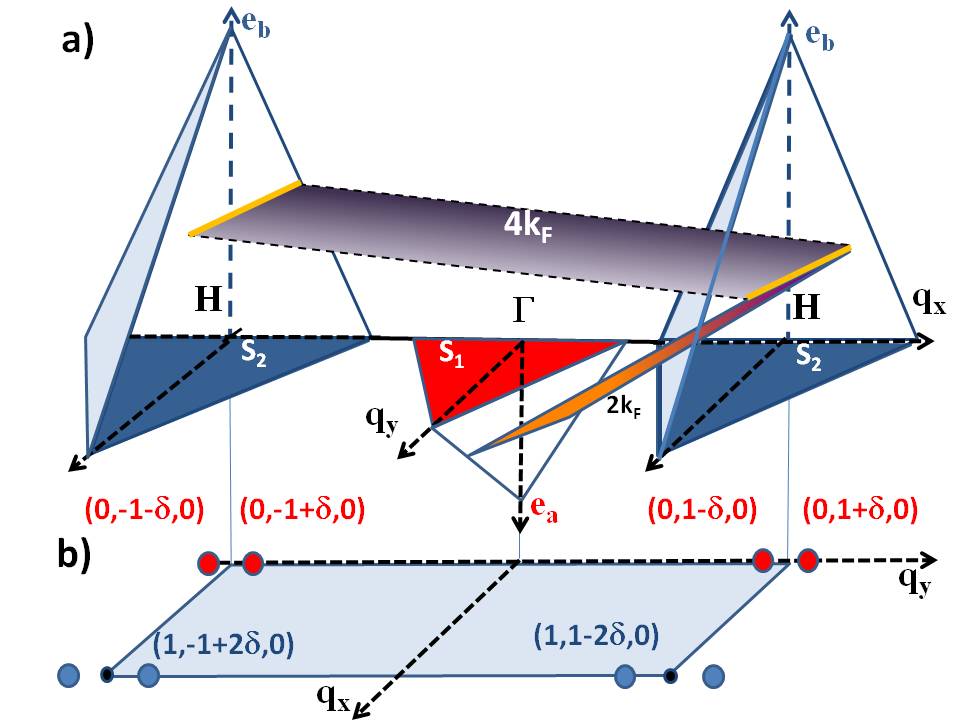}
\caption{a) Schematic Fermi surface section in the (001) plane. The red region is the electron pocket centered at $\Gamma$ and the two blue regions are the holes pockets at $H$. The partial dispersion curves are assumed to be linear, so we consider three prisms with four surfaces each. The second order nesting phenomenon is described with colored planes connecting surfaces.  b) The  (001) plane containing the satellite reflections associated to the  SDW located at $Q_S=(0,1\pm\delta,0)$ and the satellites associated to the CDW  at $Q_C=(1,1\pm2\delta,0)$.}
\label{surf_fermi} 
\end{figure}

The average behavior of
chromium does not contradict the magnetostriction scenario. Classical diffraction experiments provide spatially averaged information
 that show that both CDW and SDW appear simultaneously at the same
N\'eel temperature $T_N$ = 311K.  Micro-diffraction experiments have also shown
that their domains are highly correlated\cite{evans} and that spin and charge orders are similarly suppressed
with pressure\cite{feng}. 
Other studies, using diffraction techniques, tried to answer this issue without being able to clearly discriminate between the two scenarios\cite{mori}.

We show in this paper that a precise comparison of the {\it
 spatial fluctuations} of each modulation allows us to reconsider the origin of CDW in chromium.
Indeed, this first experiment combining coherent and simultaneous X-ray diffraction shows that  defects affect the charge order but not the spin one and hence  that spin and charge orders in chromium have the same origin: a purely electronic effect due to
the peculiar band structure of chromium.

Classical X-ray diffraction is  a powerful technique to probe the
charge and the spin modulation of chromium because  the  SDW period 
 is twice larger than the CDW one, and the two  satellite reflections can be measured independently at $Q_S=(0,1-\delta,0)$ and at  $Q_C=2Q_S=(0,2-2\delta,0)$ wave vectors. In ref \cite{hill}, the profiles of the two satellites reflections indicates that the CDW coherence length is smaller than the SDW one. 

However, a precise comparison of correlation lengths of  coexisting phases is difficult by diffraction techniques since
the probed volumes are usually not the same. In the case of chromium in \cite{hill}, the  volume  probed at $Q_S$ is much smaller that the one probed at $2Q_S$\cite{note1} which makes the comparison of  the two correlation lengths only qualitative.

We have used here a combination of  two techniques to go beyond these limitations. Coherent x-ray
diffraction has been performed to measure spatial fluctuations of the two modulations without spatial
average {\it and} simultaneous diffraction to probe
CDW and SDW in the same sample volume.

 A coherent x-ray beam has been obtained from a weakly coherent
synchrotron source by using a set of two slits. The first one is located just
after the optics to get rid of optical aberrations and the second one at 15 cm upstream of the sample. 
\begin{figure}
\begin{center}
\includegraphics[width=6cm]{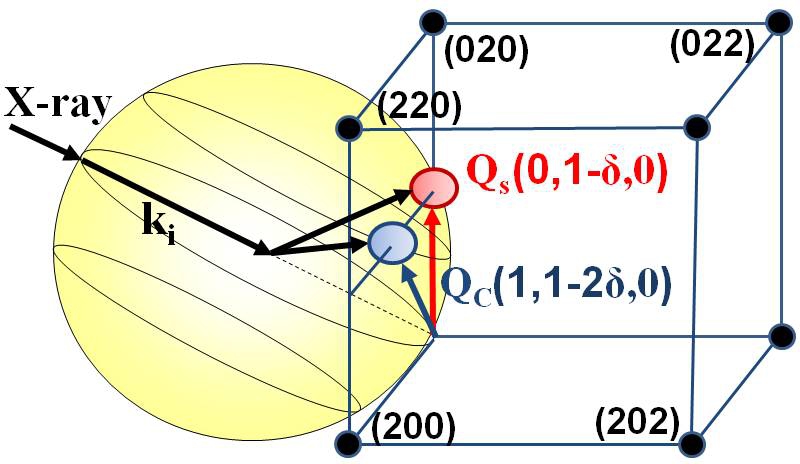}
\end{center}
\caption{Schematic drawing of the simultaneous diffraction experiment. Given an incident
 wave vector $k_i$, there exists a sample orientation for which the $Q_S$ and $Q_C$ satellite
reflections are simultaneously located on the Ewald sphere. As a consequence, both reflections can be measured only moving the detector, and
not the sample. This particular geometry ensures that the probed volume is the same when probing the two reflection satellites at $Q_c$ and $Q_s$.
\label{simultaneous} }
\end{figure}
Thanks to this setup and using  appropriate apertures, a 10{~}$\mu$m x-ray beam with
$90\%$ degree of coherence is obtained. The
experimental setup  used for coherent diffraction experiments is  described in details
in \cite{lebolloch3} and the ability of this technique to probe charge density wave systems in \cite{jacques2}.
The experiments have been performed  at the Cristal
 beam line  of  synchrotron  Soleil  and at the ID20 beam line of the ESRF using  an energy of E=5.9{~}keV,  just below the chromium $k$-edge to avoid fluorescence.
A single-$Q$ domain chromium  sample has been probed in reflection geometry with $\delta\approx0.047$  in reciprocal lattice units (r.l.u.) at T=140 K. 
 
To fulfill simultaneous diffraction conditions, the sample is placed  in such a way that
the $Q_S=(0,1-\delta,0)$ and the $Q_C=(1,1-2\delta,0)$ satellites  are simultaneously
located on the Ewald sphere.  Therefore, both reflections can be
measured by moving only the detector
and not the sample. As a consequence, the beam location on the sample's surface and the volume probed by
the beam are equal for the two satellites.
\begin{figure}
\center
\includegraphics[width=8cm]{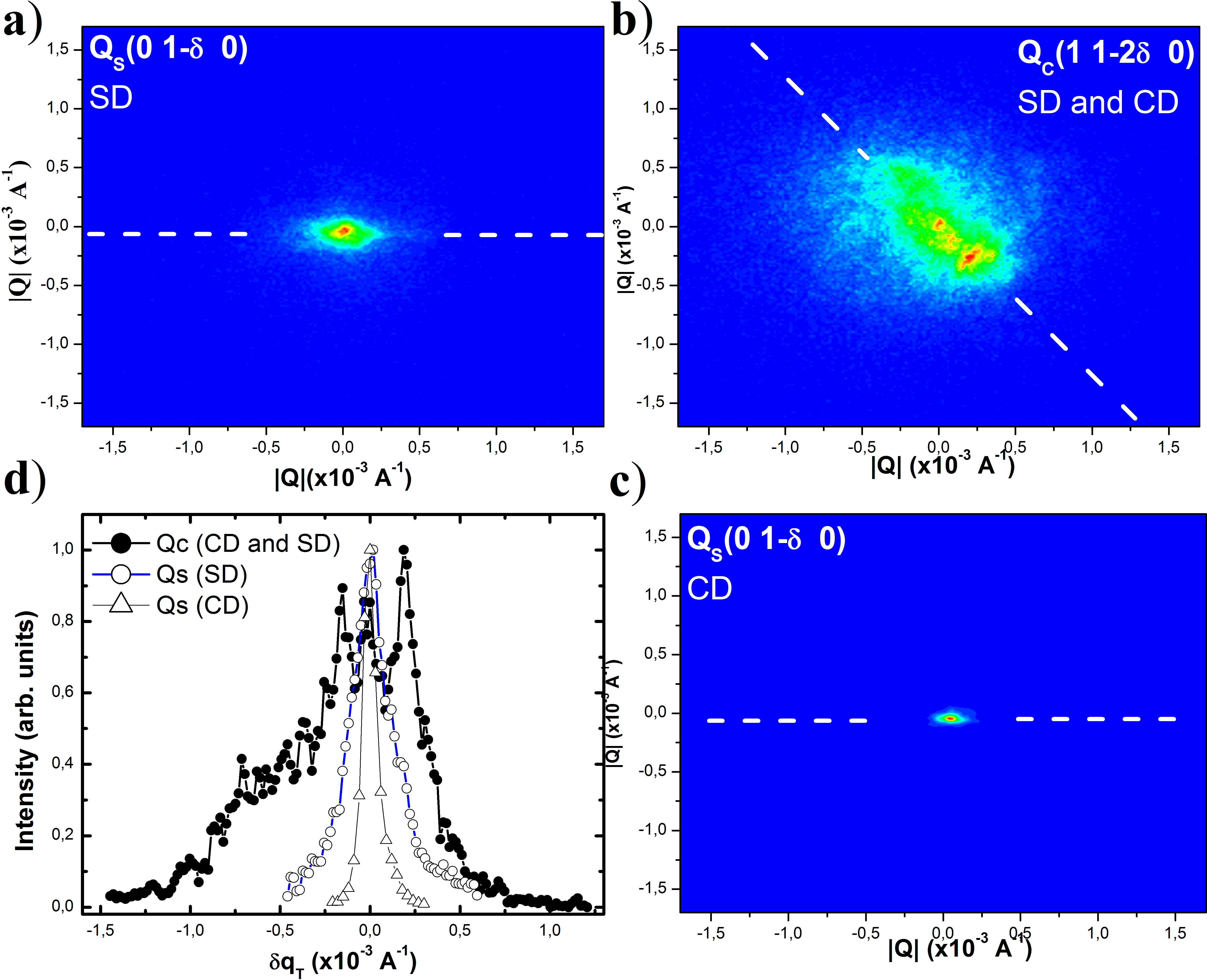}
\caption{Comparison of  SDW and CDW in chromium by using
coherent and simultaneous diffraction.  Here are the diffraction patterns through the maximum intensity of the CDW $(Q_{C}{=}(1{,}1{-}2\delta,0))$ and SDW satellites $(Q_{S}{=}(0{,}1{-}\delta,0)).$ a) Simultaneous diffraction (SD) of the SDW satellite (beam size=100$\mu m$*300$\mu m$(H)) b) Simultaneous and coherent diffraction of the CDW satellite (beam size=10$\mu m$*10$\mu m$) and c) coherent diffraction (CD) of the $Q_S$ satellite (beam size=10$\mu m$*10$\mu m$). d)  Profiles corresponding to the three patterns along the direction represented by the white dashed lines. In all cases, the
$Q_C$ satellite displays speckles, while no speckle is observed at $Q_{S}$. The maximum intensity obtained on the two reflections is 150 counts/pixel for the SDW reflection and 100 counts/pixel for the CDW reflection.}
\label{speckle} 
\end{figure}

The three diffraction patterns displayed in Fig. 2 give a clear picture  of the
SDW and  CDW states in chromium. No speckle is
observed at the satellite reflection $Q_S$ associated to the SDW.  
The pattern displays a single peak, the width of which corresponds to the
footprint of the beam at the sample surface (Fig. 3a and Fig. 3c). The
fundamental (0 1 1) reflection associated to the lattice displays also a
single peak with an equal width. In contrast, the satellite $Q_C$ associated
to the CDW is broader and displays speckles (see Fig. 3b). 

The interpretation is clear and unequivocal: the SDW in chromium displays a impressive long range order, without domains or phase
shifts, over the entire probed volume, i.e. 10$\mu\rm m\times$10$\mu\rm m\times$7$\mu$m (in depth).  Note that a single phase shift  of the magnetic order would induce a splitting of the satellite reflection in  Fig.3c)\cite{silicon}. On the contrary, in the same volume, 
the CDW displays phase shifts.  This measurement proves
that, contrary  to the charge one, the spin order is unaffected by the presence of punctual defects such as interstitials or/and
vacancies which are abundant in such a large volume. 
This conclusion differs from that of ref. \cite{naturechromium} in which the authors interpret the presence of speckles on the CDW reflection as a consequence of the presence of magnetic domain walls. In our case, the sample is single-Q and speckles are present on the CDW reflection and not on the SDW reflection. 
Thus, in our case, the presence of speckles on the CDW reflection can not be explained by the presence of antiferromagnetic domains.

This comparison between charge and spin orders is crucial to better understand
the origin of  both states. How can we explain that, despite the strong relationship between SDW and CDW in chromium, the spin correlation length remains long-range while the charge one is much smaller\cite{habibi}?  In the following, the two  scenarios
mentioned in the introduction are discussed in the light of our experimental results. 

Magnetostriction is the most frequently cited scenario to account for the presence
of CDW in chromium. The appearance of a periodic
lattice distortion within a magnetic order may reduce the total energy. In that
case, three contributions have to be considered: the anisotropy energy
($E_{\rm an}$), the exchange energy ($E_{\rm ex}$) and the elastic cost due the
periodic distortion ($E_{\rm el}$):
\[
\left\{\begin{array}{lcl} E_{\rm el}&=&\frac C2 \sum (u_n-u_{n+1})^2 \\
E_{\rm ex}&=&-\sum_{n,m} J(R_{nm})
{\langle}J_n{\rangle}{\langle}J_m{\rangle}\\
E_{\rm an}&=&-\sum_n K_2(x_{n+1}-x_{n-1}) \mu^2_n \end{array}\right.\ 
\]
where $\mu_n$ is the magnetic momentum of the $n^{\rm
th}$ atom, $u_n$  the atomic displacement with respect to the non-magnetic structure, $K_2$ depends of the distance between planes $n+1$ and $n-1$, $C$ the force constant
 and $J_n$ the kinetic momentum.  In chromium,  the lowest total energy is found when  the CDW period is  twice shorter than 
that of  SDW, in agreement with experiments\cite{nourtier}.  The main  point here is that the energy  minimization
gives a close relationship between CDW and SDW:   spatial fluctuations of the first  lead to spatial fluctuations of the second and reciprocally. 
From this point of view, the magnetostriction scenario seems hardly compatible
with our experimental data displayed in Fig. 3.

In the following, we consider the second scenario introduced by Young \& Sokoloff in \cite{youngsokoloff} 
based on a three-band model. We show that this model, in the presence of punctual defects,  is in agreement with our measurement. In
this approach, the CDW is the second harmonic of  SDW. More generally,
these authors have shown that  odd harmonics are spin orders and 
 even ones are charge orders. 

 In this purpose,  inelastic nesting between bands has been considered as in \cite{fishman} and  linear dispersion curves  for electrons ($\varepsilon_a(q_x,q_y)$) and holes ($\varepsilon_b(q_x,q_y)$)  have been assumed. We thus have to consider three prisms and connect the four surfaces of each of them. The model is adjusted to the case of chromium\cite{RefJ3}; the size of the electron ($s_1$) and hole ($s_2$) pockets  is  such as $\delta=s_2-s_1$. The two Fermi velocities ($v_e$ and $v_t$) have also been  extracted from experimental data. The three corresponding prisms with  maximum energies $e_a=-s_1v_e$ and $e_b=s_2v_t$ are drawn of Fig. 1.a and in Fig. 4a.
For a  strict coexistence of both phases such as measured in chromium, a simultaneous nesting is authorized between the electron and the hole pockets with $|Q_{2kF}|=1\pm\delta$ {\it and} between the two hole pockets  at $|Q_{4kF}|=2\pm2\delta$. 
The three-band hamiltonian reads:
\[
{\cal H}=\pmatrix{\varepsilon_a-i\omega&\Delta_s&\Delta_s\cr
\Delta_s&\varepsilon_b-e_0\delta-i\omega&\Delta_c\cr 
\Delta_s&\Delta_c&\varepsilon_b+e_0\delta-i\omega\cr}
\]
where $\Delta_s$ is the SDW order parameter and  $\Delta_c$ the CDW order
parameter.  The linear temperature dependence of  $\delta$ has been
extracted from \cite{fawcett}.  The Green function can be extracted from 
${\cal G}(i\omega-{\cal H)}=I$ following the standard BCS equation:
\[
\pmatrix{\varepsilon_a&\Delta_s&\Delta_s\cr
\Delta_s&\varepsilon_b&\Delta_c\cr 
\Delta_s&\Delta_c&\varepsilon_b\cr}
\pmatrix{{\cal G}_a&{\cal F}_s&{\cal F}_s\cr
{\cal F}^\dag_s&{\cal G}_b&{\cal F}_c\cr 
{\cal F}^\dag_s&{\cal F}^\dag_c&{\cal G}_b\cr}=
\pmatrix{1&0&0\cr0&1&0\cr0&0&1\cr}\ .
\]
Making the summation of ${\cal F}_s$ and ${\cal F}_c$ over all Matsubara
frequencies, one finds two simultaneous gap equations
\begin{eqnarray*}
1\over{g_s}&=&\int dq_xdq_y\Bigg(
{(\varepsilon_b-x_0+e_0\delta-\Delta_c)
\tanh({x_0\over2T})\over2(x_0-x_1)(x_0-x_2)}\\
&&+{(\varepsilon_b-x_1+e_0\delta-\Delta_c)
\tanh({x_1\over2T})\over2(x_0-x_1)(x_2-x_1)}\\
&&+{(\varepsilon_b-x_2+e_0\delta-\Delta_c)
\tanh({x_2\over2T})\over2(x_0-x_2)(x_1-x_2)}\Bigg)\\
\hbox{and}\cr
1\over{g_c}&=&\int dq_xdq_y\Bigg(
{(\varepsilon_a-x_0-\Delta_s^2/\Delta_c)
\tanh({x_0\over2T})\over2(x_0-x_1)(x_0-x_2)}\\
&&+{(\varepsilon_a-x_1-\Delta_s^2/\Delta_c)
\tanh({x_1\over2T})\over2(x_0-x_1)(x_2-x_1)}\\
&&+{(\varepsilon_a-x_2-\Delta_s^2/\Delta_c)
\tanh({x_2\over2T})\over2(x_0-x_2)(x_1-x_2)}\Bigg)
\end{eqnarray*}
where $x_0$, $x_1$ and $x_2$ are the three zeros of $\det({\cal H})$,
$e_0=e_b-e_a$,  $g_s$  is the coupling constant between electron and hole and $g_c$ the coupling between holes.  We are
looking for simultaneous solutions of the two gap equations corresponding to 
a stable thermodynamical phase. Solutions of each gap equation separately correspond to
single (spin or charge) phases, when the other order parameter is
suppressed.
\begin{figure}
\center
\includegraphics[width=8cm]{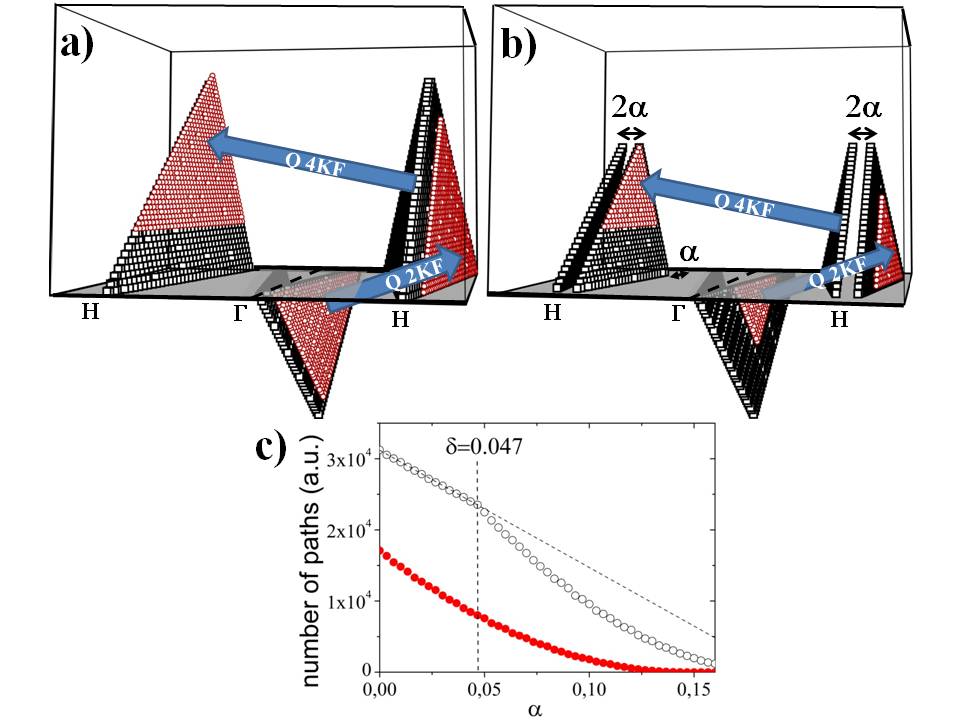}
\caption{Authorized paths between the electron  and  hole pockets $(Q_{2kF}\!)$ and the two hole pockets $(Q_{4kF}\!)$. Red circles are in-\penalty-100000 volved in the inelastic process. Black squares do not fulfill the necessary conditions and are not involved in the nesting:  a) in the perfect case and b) in the case with defects where states at the border of the Brillouin zone in two $2\alpha$ width strips are removed from the process. c) The number of authorized paths between the electron and hole pockets (open circles) and between the two hole pockets (full circles) versus $\alpha$. }
\end{figure}
The integration in the $(q_x,q_y)$ plane taking into account all  nesting
processes corresponds to the summation over the 4 surfaces of each prism. 
Considering only one surface out of four, the  process is described in Fig. 4a where the points fulfilling simultaneously both conditions ($|Q_{2kF}|=1\pm\delta$ {\it and} $|Q_{4kF}|=2\pm2\delta$ along
the [010] direction only) are represented with red circles. Note that the electron surface, which is smaller,
constrains the hole surface and restricts the total number of nested points.  
Within this model, we correctly reproduce the mixed
state in chromium and  the Néel temperature of T=311K. A paper detailing theoretical aspects will be published elsewhere.

In this approach,  CDW is the second harmonic of  SDW and will be more sensitive to punctual defects in agreement with our measurement.
One can simply understand this greater sensibility by considering the influence of punctual defects on the three prims of Fig. 1. 
Indeed, similarly to  phonons in the presence of punctual defects\cite{note2}, we consider that  the dispersion curves will be mainly affected at the border of Brillouin zone.
Therefore, only the two hole pockets centered at  points $H$ will be affected and not the electron one at $\Gamma$.  To take into account this effect, we remove from  nesting processes the states located in two  $2\alpha$ wide stripes (one along $q_y$ and one along $q_x$) centered at  points $H$ (see Fig. 4b). As a consequence, the contribution of the second order nesting in the total energy is strongly reduced with respect to the contribution of the  first order one. This
statement becomes clear if  the number of authorized paths between bands is taken into account: the number of paths between the two hole pockets decreases much more sharply than the number of paths between the electron and hole pockets for increasing $\alpha$. Since the two holes areas are affected by punctual defects, the first one decreases as $\alpha^2$, while the second one decreases as linearly for $\alpha<\delta$ (see Fig. 4c). 

In conclusion, an original experiment coupling coherent x-rays and simultaneous diffraction has been performed to precisely measure correlation lengths in chromium. The SDW does not display any dislocation over several micrometers in the sample while the CDW one displays many speckles. This observation is a clear misstatement of the usually accepted magnetostriction theory. We explain these experimental features using the Young and Sokoloff model where the CDW and the SDW are coming from the same phase with the CDW being a second harmonics of the SDW. This second harmonic theory and the peculiar band structure of chromium makes the CDW more sensitive to punctual defects, in agreement with our measurement.
From this approach, we predict that  in the limit case, in the presence of many uncorrelated defects, a new phase of chromium should stabilize  SDW  with no CDW. In this framework, the $Q_{4kF}$ wave vector is not associated to a simple strain wave but to a true incommensurate CDW linked to a periodic lattice distortion. From this point of view, the CDW in chromium should slide under an external current.


\begin{thebibliography}{00}
\bibitem{note1} The x-ray penetration is $7\mu m$  in depth at $Q$ and $14 \mu m$ at $2Q$ at the incident energy E=5.9{~}KeV.
\end{thebibliography}
\end{document}